\documentclass{webofc}
\usepackage[varg]{txfonts}   
\usepackage{natbib}
\usepackage{lineno}
\usepackage[utf8]{inputenc}
\usepackage{float}
\usepackage{booktabs} 
\usepackage{placeins} 
\usepackage{comment} 
\usepackage{subcaption}
\usepackage{graphicx}
\usepackage{adjustbox}
\begin{document}
\title{Advancements in Computing and Simulation Techniques for the HIBEAM-NNBAR Experiment}
%
%
\author{\firstname{First author} \lastname{First author}\inst{1,3}\fnsep\thanks{\email{Mail address for first
    author}} \and
        \firstname{Second author} \lastname{Second author}\inst{2}\fnsep\thanks{\email{Mail address for second
             author if necessary}} \and
        \firstname{Third author} \lastname{Third author}\inst{3}\fnsep\thanks{\email{Mail address for last
             author if necessary}}
}

\institute{Insert the first address here 
\and
           the second here 
\and
           Last address
          }

\author{
\firstname{Bernhard} \lastname{Meirose}\inst{1,2}\fnsep\thanks{\email{bernhard.meirose@fysik.lu.se.}} \and
\firstname{Jorge} \lastname{Amaral}\inst{6} \and
\firstname{Alexander} \lastname{Burgman}\inst{3} \and
\firstname{Matthias} \lastname{Holl}\inst{1} \and
\firstname{Ernesto} \lastname{Kemp}\inst{7} \and
\firstname{Adam} \lastname{Kozela}\inst{9} \and
\firstname{David} \lastname{Milstead}\inst{3} \and
\firstname{André} \lastname{Nepomuceno}\inst{4} \and     
\firstname{Anders} \lastname{Oskarsson}\inst{1} \and 
\firstname{Krzysztof} \lastname{Pysz}\inst{9} \and
\firstname{Valentina} \lastname{Santoro}\inst{1,5} \and 
\firstname{Tiago} \lastname{Quirino}\inst{6} \and
\firstname{Blahoslav} \lastname{Rataj}\inst{1} \and
\firstname{Gabriel} \lastname{Silva}\inst{6} \and 
\firstname{Samuel} \lastname{Silverstein}\inst{3} \and
\firstname{Magnus} \lastname{Wolke}\inst{8} \and
\firstname{Lucas} \lastname{Åstrand}\inst{1}
       }
\institute{
Fysiska institutionen, Lunds universitet, 221 00, Lund, Sweden \and  
Institutionen f{\"o}r Fysik, Chalmers Tekniska H\"{o}gskola, Sweden \and 
Department of Physics, Stockholm University, 106 91, Stockholm, Sweden \and 
Departamento de Ciências da Natureza, Universidade Federal Fluminense, 28895-532, Rio das Ostras, RJ, Brazil  \and 
European Spallation Source ERIC, 225 92, Lund, Sweden\and  
Faculdade de Engenharia, Universidade do Estado do Rio de Janeiro, 20950-000, Rio de Janeiro, RJ, Brazil \and 
Instituto de Física \textit{Gleb Wataghin}, Universidade Estadual de Campinas - UNICAMP, 13083-859, Campinas, SP, Brazil \and 
Department of Physics and Astronomy, Uppsala University, Uppsala, Sweden \and 
Institute of Nuclear Physics Polish Academy of Sciences, PL-31342
Krakow, Poland 
          }

\abstract{The HIBEAM-NNBAR program is a proposed two-stage experiment at the European Spallation Source focusing on searches for baryon number violation processes as well as ultralight dark matter. This paper presents recent advancements in computing and simulation, including machine learning for event selection, fast parametric simulations for detector studies, and detailed modeling of the time projection chamber and readout electronics.
} 

\maketitle

\FloatBarrier
\section{Introduction}
\label{sec:intro}
\setlength{\abovecaptionskip}{1.0mm} 
\setlength{\parskip}{0pt}
The HIBEAM-NNBAR experiment \citep{Addazi:2020nlz} is a multidisciplinary, two-stage program at the European Spallation Source (ESS)~\cite{Garoby_2017}. The first stage, HIBEAM \cite{Santoro:2023izd}, includes high-sensitivity searches for neutron oscillations, sterile neutrons, and axions, as well as searches for exotic neutron decays. The second stage, NNBAR \cite{Santoro:2024lvc}, will have a discovery potential three orders of magnitude higher than the last neutron-antineutron oscillation search conducted in the 1990s at the Institut Laue-Langevin \cite{Baldo-Ceolin:1994hzw}. This paper summarizes advancements in computing and simulation for the HIBEAM-NNBAR program, building on the developments presented in our previous paper~\cite{Barrow:2021deh}.

The paper is organized as follows. Section~\ref{ML} describes some of the machine learning techniques implemented in HIBEAM, followed by a discussion on the use of fast parametric simulations in Section~\ref{fast}. Section~\ref{sec:TPC} focuses on simulations of the time projection chamber (TPC), while Section~\ref{sec:readout} describes the simulation framework designed to replicate detector readout behavior. A brief summary is presented in Section~\ref{summary}.


\FloatBarrier
\section{Machine learning}
\label{ML}

Machine learning (ML), a branch of artificial intelligence (AI), enables computers to learn from data without explicit programming. It is particularly effective for problems lacking deterministic solutions, where algorithms identify patterns in data. In HIBEAM-NNBAR, ML enhances neutron-antineutron annihilation event selection, improving signal discrimination from background processes.

The HIBEAM simulation dataset, used for training and evaluation, contains 369,569 events. The signal consists of neutrons converting into antineutrons and annihilating within a carbon nucleus, producing multi-pion events. Background events, primarily cosmic rays, were generated using the CRY library~\cite{CRY} and simulated interacting with the annihilation detector. Each event is represented by 49 high-level variables, including \textit{TPC\_tracks} (number of TPC tracks), \textit{Scint\_eDep} (scintillator energy deposition), and \textit{Inv\_mass} (invariant mass). The dataset is split into training (40\%), validation (10\%), and test (50\%) sets. Models tested include Logistic Regression, Decision Trees, Random Forests, Extreme Gradient Boosting (XGBoost), and Light Gradient Boosting (LGBM). These were implemented in Python using the libraries scikit-learn (version 1.2.2),  XGBoost (version 2.0.3), and LightGBM (version 4.0.5), respectively.

HIBEAM-NNBAR requires 100\% background rejection. To achieve this, we developed a custom metric penalizing false positives:

\begin{equation}
    C_{\text{metric}} = \frac{FN + (\beta-1) \cdot FP}{\beta}
\end{equation}
where \textit{FN} and \textit{FP} are false negatives and false positives, respectively, and $\beta=1000$. The model, trained with this metric, is optimized using the validation set to determine the optimal threshold. Hyperparameter tuning is performed using the OPTUNA framework~\cite{akiba2019}. It presents a straightforward interface that can be used with the mentioned models to find he best hyperparameters. 

\begin{table}[htbp]
\caption{Signal/background separation results in the test set.}\label{tab:results}
\centering
\begin{tabular}{p{3.5cm} c c c c c} 
\toprule
Model & Metric & Variable & Threshold & Rejection & Efficiency\\
\midrule
Random Forest (Optuna) & Custom & 49  & 0.933 & 100 & 95.81 \\ 
XGB (Optuna)           & Custom & 49  & 0.999 & 100 & 98.71\\
LGBM (Optuna)          & Custom & 49  & 0.999 & 100 & 98.69  \\
LGBM (Optuna) + Corr.  & Custom & 26 & 0.997 & 100 &  98.00 \\
LGBM (Optuna) + PCA    & Custom & 17 & 0.999 & 100 & 74.00 \\
\bottomrule
\end{tabular}
\end{table}

Table~\ref{tab:results} summarizes the results. Random Forest, XGB, and LGBM achieved 100\% background rejection with high efficiency. To simplify the analysis, we applied correlation selection, reducing the number of variables to 26 while maintaining 100\% rejection and 98\% efficiency. Principal Component Analysis (PCA) further reduced variables to 14, but efficiency dropped to 74\%.

These results demonstrate that 100\% background rejection is achievable with ML. Current efforts focus on minimizing the number of input variables and applying explainable AI techniques~\cite{samek2019explainable} to derive interpretable insights, potentially offering new perspectives on standard cut-based selections.

\section{Fast Simulation}
\label{fast}

Full Monte Carlo-based simulations are computationally expensive. In early detector design studies, such detail may be unnecessary. Instead, a faster approach, with reduced complexity, improves processing speed.

Fast simulation techniques, widely used in high-energy physics, replace step-by-step particle interactions with parameterized detector responses. This work develops fast simulations for the HIBEAM-NNBAR experiment, particularly for the electromagnetic calorimeter.

\FloatBarrier
\subsection*{Fast Simulation of the HIBEAM Calorimeter}

This approach was used to develop a fast simulation for the NNBAR electromagnetic and hadronic calorimeters, as well as for a potential HIBEAM electromagnetic calorimeter. The NNBAR and HIBEAM electromagnetic calorimeters are made of lead glass, while the NNBAR hadronic calorimeter consists of a stack of 10 scintillator slats. In this section, we focus on the fast simulation of the HIBEAM electromagnetic calorimeter.

The WASA Scintillator Electromagnetic Calorimeter (SEC) was part of an experiment aimed at investigating light meson production in light ion collisions and rare $\eta$ meson decays at the CELSIUS storage ring in Uppsala~\cite{Bargholtz_2008}. The SEC is a strong candidate for the HIBEAM electromagnetic calorimeter.  

To study the performance of the WASA SEC in reconstructing neutral pions, charged pions, and protons, a fast detector simulation was developed. For photons, the parametrization of the signal of interest—energy deposits—was constructed based on the detector's energy resolution for a two-photon signal, as given in reference~\cite{Bargholtz_2008}:  

\begin{equation}  
\frac{\sigma_E}{E} = \frac{0.05}{\sqrt{E [\text{GeV}]}}  
\end{equation}  

This parametrization was implemented in Geant4~\cite{GEANT4:2002zbu}. A simplified version of the SEC was used, where the actual calorimeter geometry was modeled as a hollow cylinder with an inner radius of 32.5 cm, an outer radius of 63.5 cm, and a length of 109 cm. When an incoming particle (photon or electron) reaches the calorimeter, the deposited energy is determined through Gaussian smearing of the particle's kinetic energy, with the Gaussian width set by the detector resolution.  

The code was validated by simulating the decay of $\eta$ mesons into photons and neutral pions (which subsequently decay into photons) and comparing the fast detector simulation results with real data recorded by the WASA calorimeter. We found good agreement between the fast simulation and real data. Additionally, a fast simulation of photon-pair detection from the decay of neutral pions—a key signal for neutron annihilation studies—was performed. Note that at this stage, fast simulation smears photon energies only, without altering their positions or simulating shower development. Thus, shower overlaps are not considered, and photons are well separated by construction.

\begin{figure}[!htbp]
  \centering
  \begin{subfigure}{0.46\textwidth}
    \centering
    \adjustbox{valign=t}{\includegraphics[width=\textwidth]{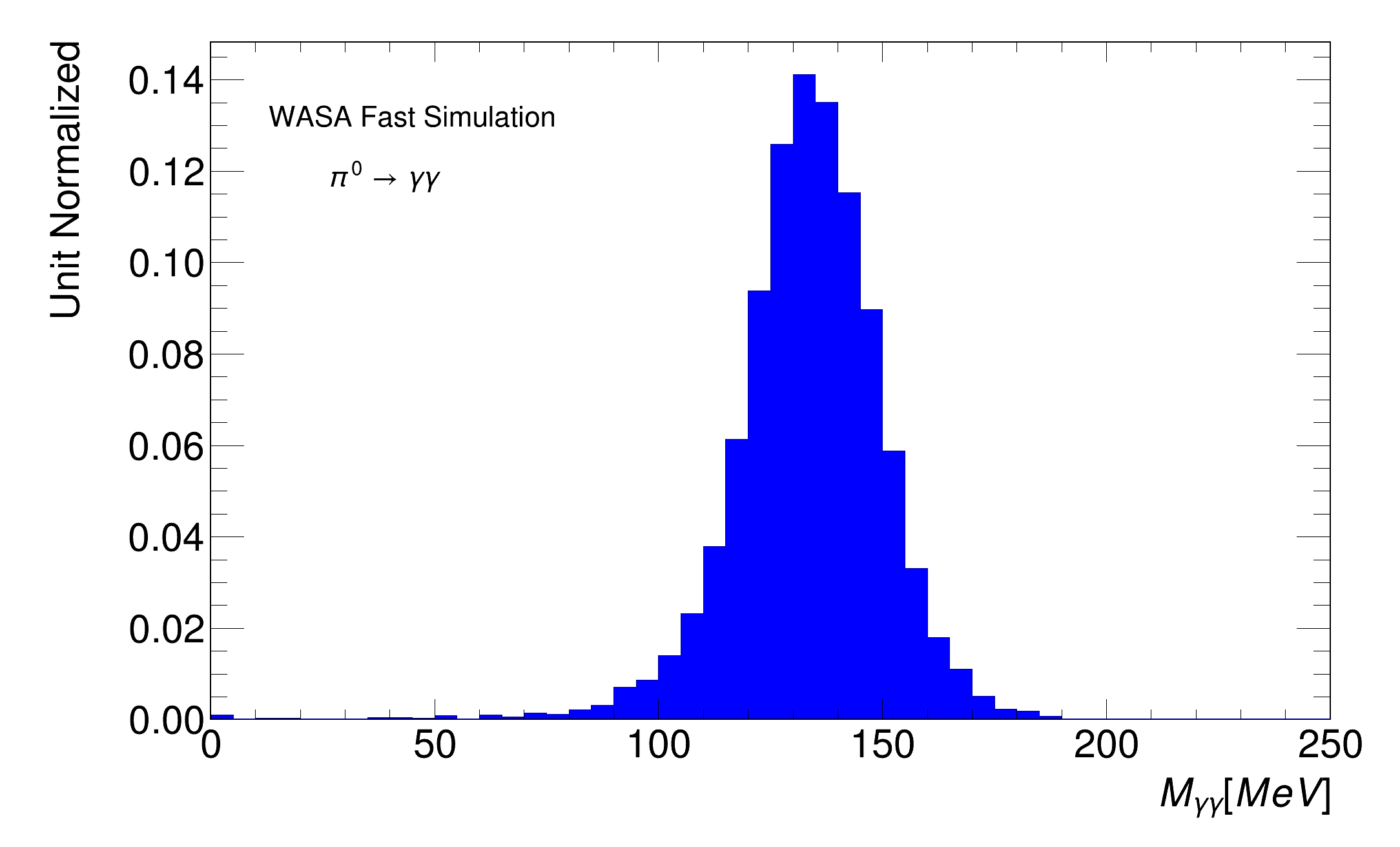}}
    \subcaption{}
    \label{fast1a}
  \end{subfigure}
  \begin{subfigure}{0.40\textwidth}
    \centering
    \adjustbox{valign=t}{\includegraphics[width=\textwidth]{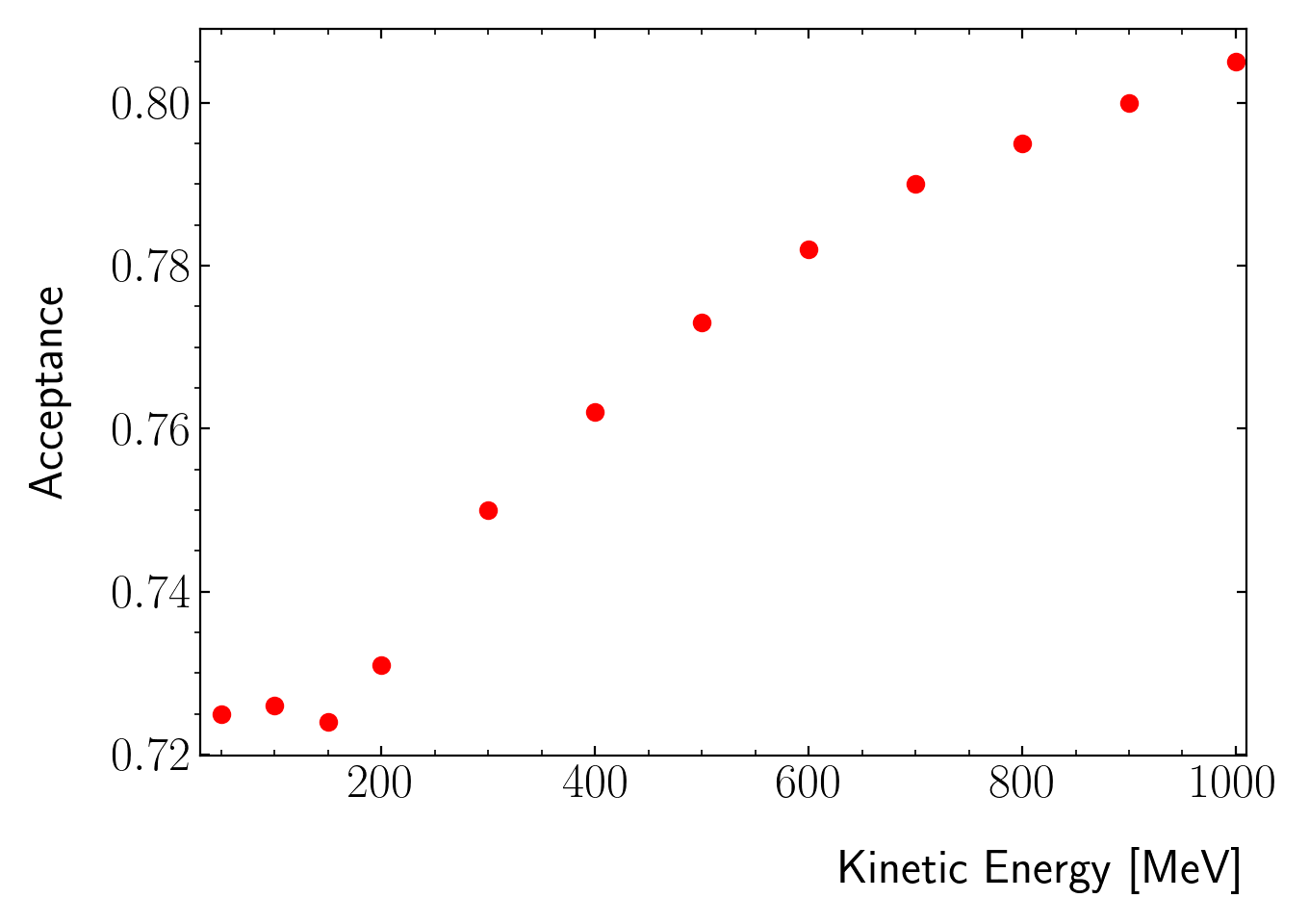}}
    \subcaption{}
    \label{fast1b}
  \end{subfigure}
  \caption{(a) Reconstructed $\pi^0$ invariant mass and (b) SEC geometrical acceptance.}
  \label{fast1}
\end{figure}

Figure~\ref{fast1a} shows the reconstructed neutral pion mass from the fast simulation of a photon pair detected by the calorimeter, originating from the decay of neutral pions with kinetic energies of 200 MeV. Figure~\ref{fast1b} illustrates the calorimeter’s geometrical acceptance as a function of $\pi^0$ kinetic energy. As shown, for energies around 150 MeV—the typical energy of $\pi^0$ from annihilation—the acceptance is approximately 73\%.

\begin{figure}[!htbp]
  \centering
  \begin{subfigure}{0.45\textwidth}
    \centering
    \adjustbox{valign=t}{\includegraphics[width=\textwidth]{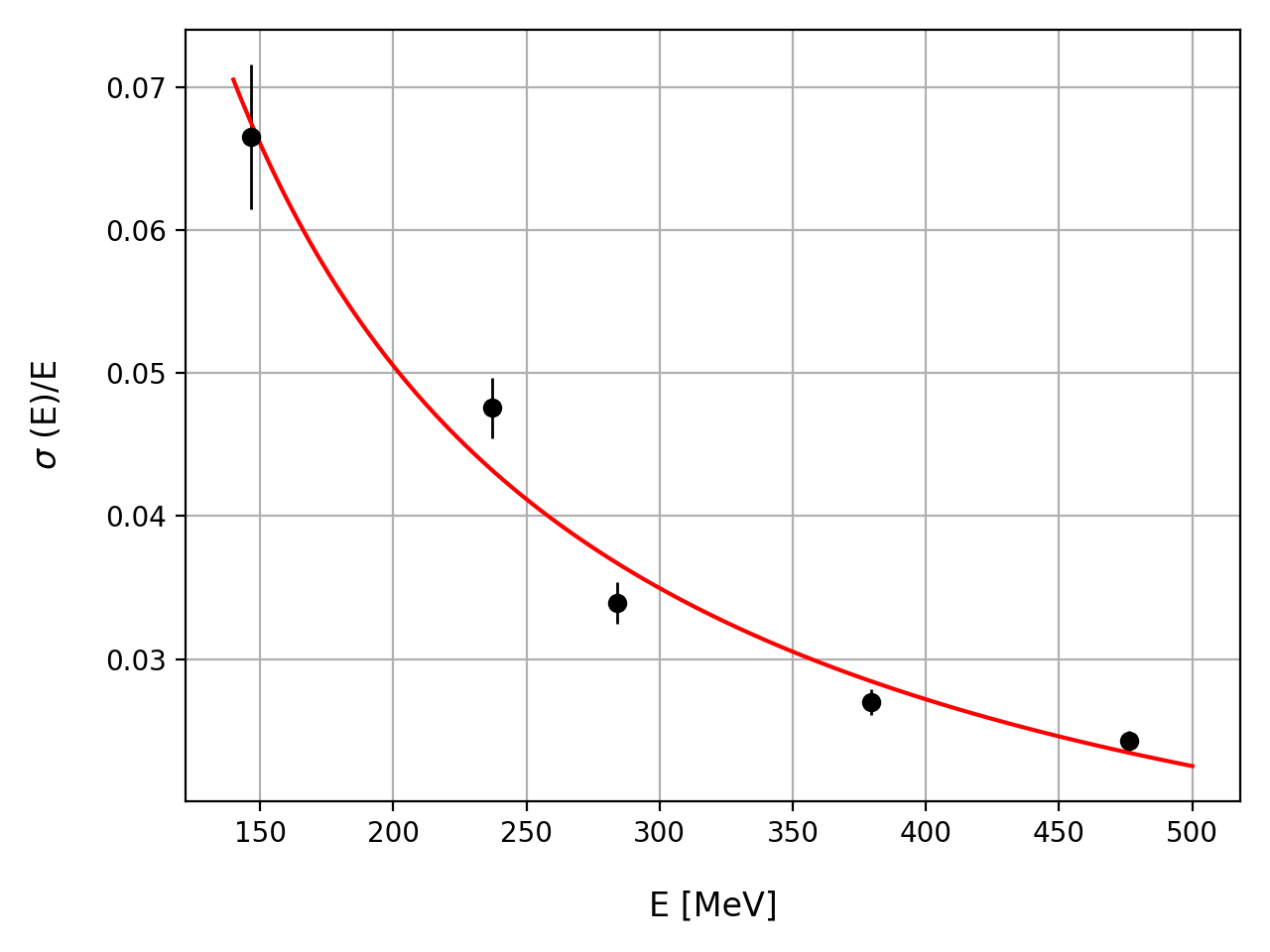}}
    \subcaption{}
    \label{fast2}
  \end{subfigure}
  \begin{subfigure}{0.53\textwidth}
    \centering
    \adjustbox{valign=t}{\includegraphics[width=\textwidth]{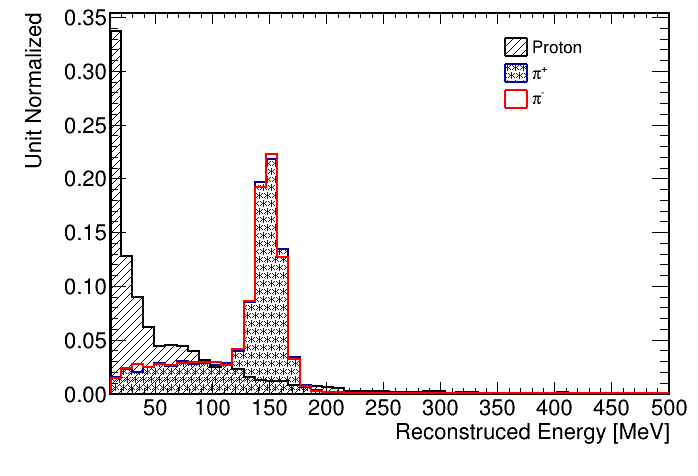}}
    \subcaption{}
    \label{fast3}
  \end{subfigure}
  \caption{(a) Energy loss resolution for charged pions in a CsI crystal, with fitted curve, and (b) fast simulation of annihilation signal in the HIBEAM electromagnetic calorimeter.}
  \label{fast4}
\end{figure}

For the charged pion response, the data used to derive the parametrization were obtained from an experiment using a detector array of seven pure CsI crystals, with dimensions similar to those of the SEC crystal, located at the proton synchrotron at KEK~\cite{Yamazaki:1997zv}. CsI crystals effectively stop charged pions with kinetic energies up to $\sim$ 150 MeV. For kinetic energies above $\sim$200 MeV, the pions punch through the detector unless nuclear reactions occur, resulting in an energy loss peak at around 140 MeV, which is nearly independent of the incident kinetic energy. The spectrum for $\sim$190 MeV pions includes contributions from both stopped and punched-through pions and is treated as a special case in the parametrization.         
\newline
A Gaussian function is fitted to the pion energy distributions in the crystal for different initial kinetic energies, and the fitted Gaussian distribution widths are used to determine the resolution. Figure~\ref{fast2} shows the resolution obtained for the pion kinetic energies available in the data, along with a fitted resolution function.

The parametrization for charged pion energy loss is given by

\begin{equation} 
\frac{\sigma_E}{E} = \frac{9.3}{E [\text{MeV}]}. 
\end{equation}

For the special case (kinetic energy $\sim$ 190 MeV), the resolution is $\sigma_E / E = 0.16$. 
\newline   

A similar approach was used to simulate the detector response to protons, using data from~\cite{MERCHEZ1989133}. The resolution model is given by

\begin{equation} 
\frac{\sigma_E}{E} = \frac{a}{\sqrt{E \,[\text{GeV}]}} \,+\, b, 
\end{equation}
with fitted parameters $a = 0.00068$ and $b = 0.0036$.
\newline
Figure~\ref{fast3} shows a fast simulation of charged pions and protons expected from the annihilation signal in the HIBEAM calorimeter. The peak observed at around 150 MeV reflects the fact that pions with higher kinetic energies punch through the detector and contribute to an energy peak at $\sim$ 150 MeV. 

\section{TPC Simulation} 
\label{sec:TPC}
A TPC is a detector that reconstructs charged particle tracks in a gaseous volume. Fast readout electronics enable 3D tracking in high track-density environments. As a particle traverses the gas, it ionizes atoms along its path. Under a uniform electric field, the ionization electrons drift toward a segmented anode, forming a 2D (XY) projection. The third coordinate (Z) is determined from the drift time, which depends on the gas properties.

The planned TPC features charge amplifier GEM detectors in the XY plane, followed by charge collection pads. An electric field in the induction region reduces electron cloud diffusion, enhancing spatial resolution. The collected charge's impact positions (XY) and known drift time enable full track reconstruction.

Three key studies are required for TPC development: (1) Computational simulations using Garfield++~\cite{schindler_heinrich_garfield_2010} to assess resolution, (2) Optimization of readout pad geometry, considering zigzag pads for improved resolution, and (3) Readout electronics development.

\begin{figure}[htb!]
  \begin{center}
    \includegraphics[width=0.40\textwidth]{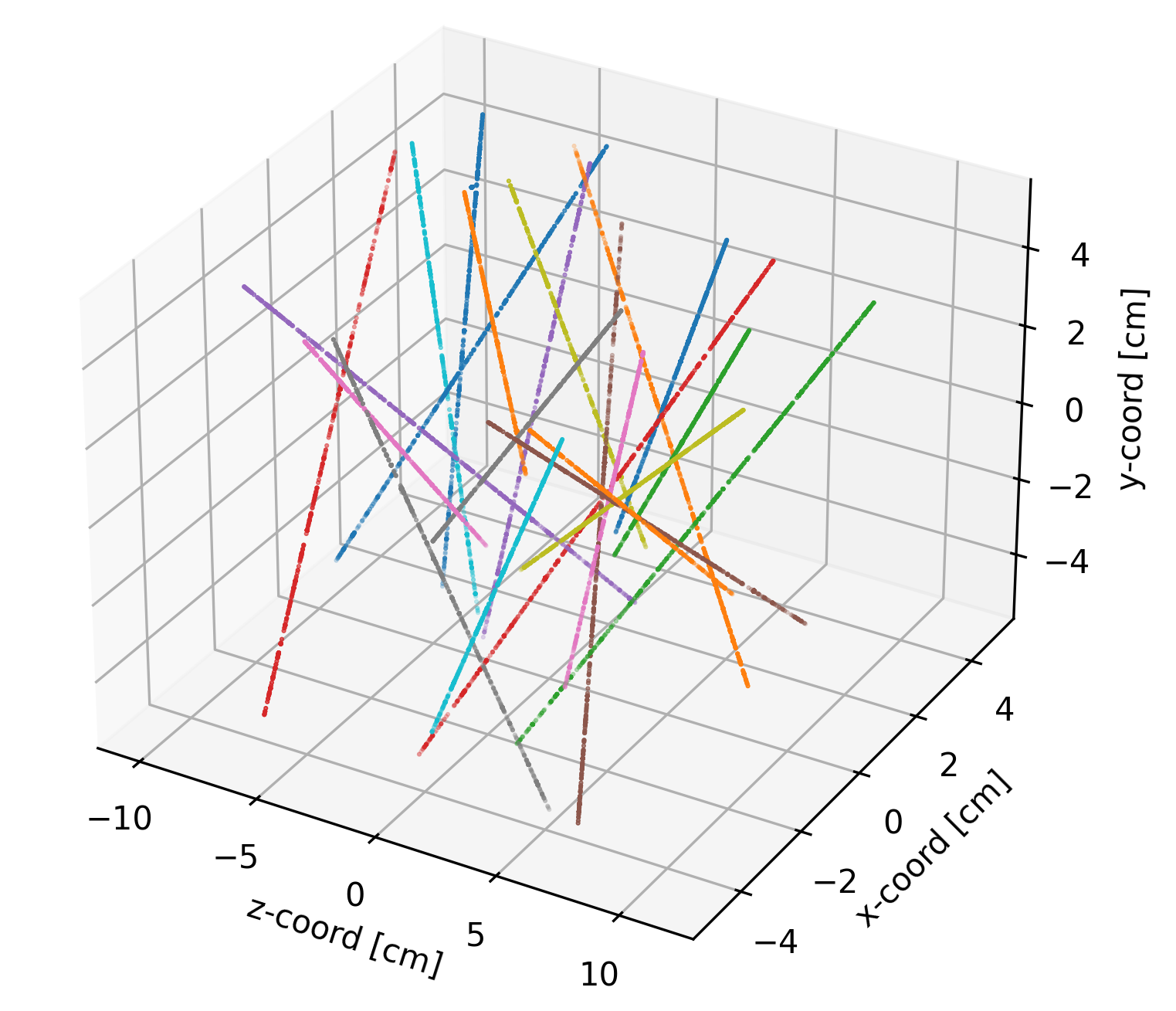}
  \end{center}
  \caption{Simulated muon trajectories in the TPC.}
  \label{traj}
\end{figure}

The simulation framework integrates Matlab\copyright~\cite{MATLAB}, COMSOL\copyright~\cite{COMSOL}, and Garfield++~\cite{Pfeiffer:2018yam}. Matlab\copyright ~ generates TPC geometries, which COMSOL\copyright ~ processes for mesh generation and electric field calculations. The output feeds into Garfield++, which simulates particle interactions with gas. The setup tested includes a $23.1 \times 10.0 \times 10.0$ cm$^3$ TPC with three $10 \times 10$ cm$^2$ GEMs, an 80\% Ar/20\% CO$_2$ gas mixture, and a 287.3 V/cm field. A sample of 21 muons (1 GeV) is simulated in random directions. The framework allows rapid configuration changes via Matlab scripting. Garfield++ outputs ROOT files containing electron drift data for analysis. Using parallel computing, multiple particles are simulated concurrently. Figure~\ref{traj} displays reconstructed muon trajectories.

\begin{figure}[!htbp]
  \begin{center}
    \begin{minipage}{0.25\textwidth}
      \includegraphics[width=\textwidth]{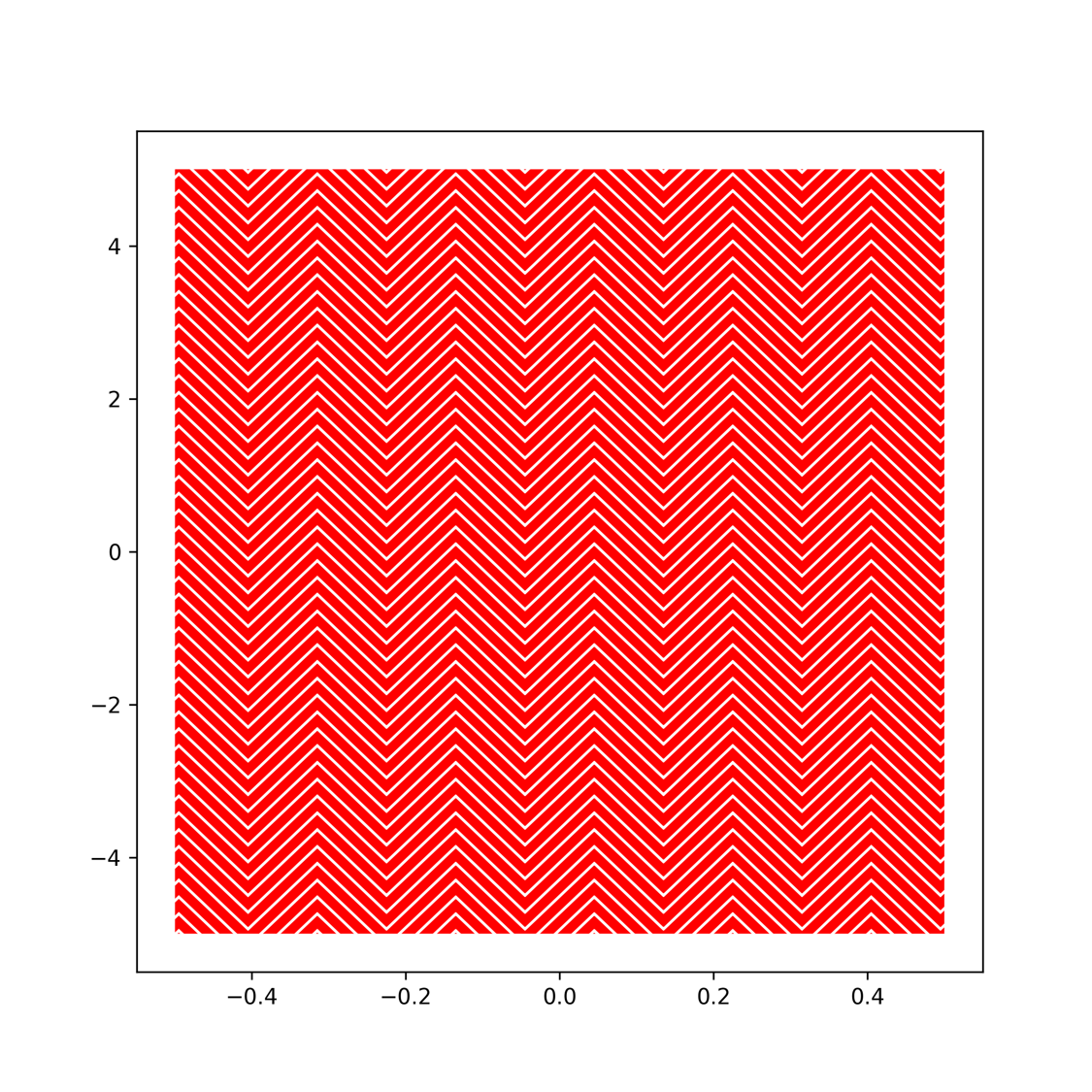}
    \end{minipage}
    \begin{minipage}{0.25\textwidth}
      \includegraphics[width=\textwidth]{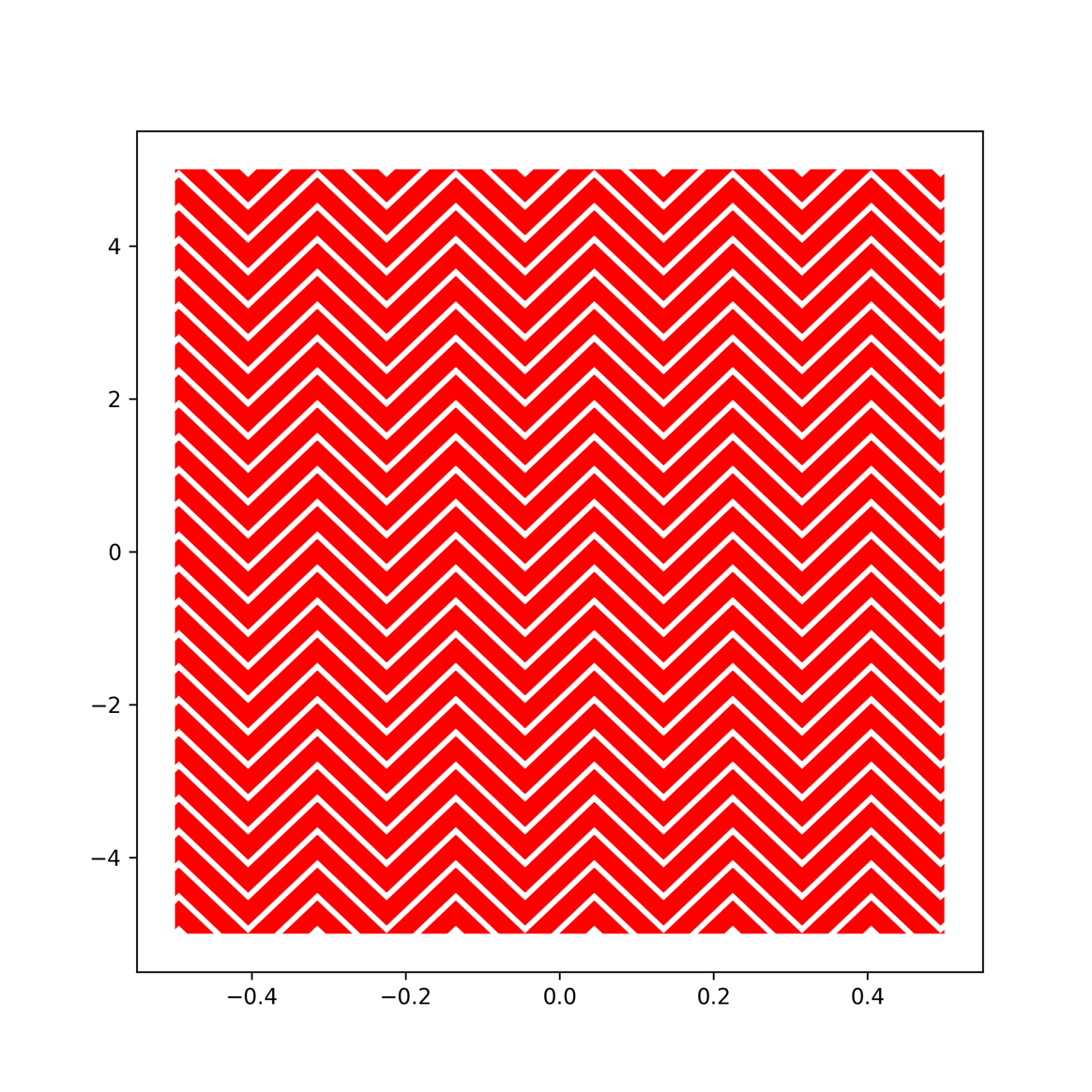}
    \end{minipage}
     \begin{minipage}{0.25\textwidth}
      \includegraphics[width=\textwidth]{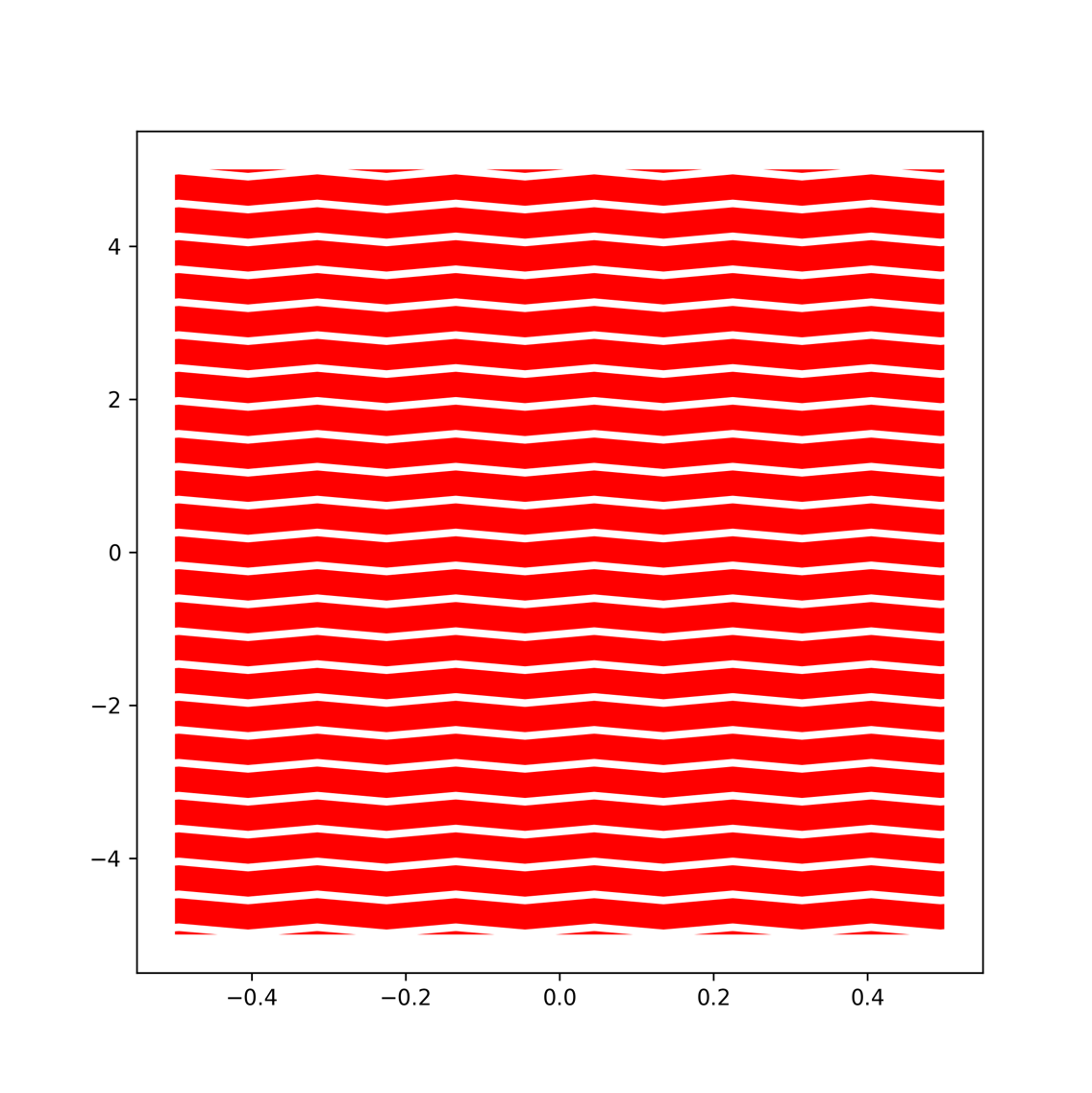}
    \end{minipage}
  \end{center}
  \caption{Zigzag pad designs: Double pads (left), Prototype (center), Double angle (right).}
  \label{zigzag}
\end{figure}

\begin{figure}[htb!]   
  \begin{center}
    \includegraphics[width=0.65\textwidth]{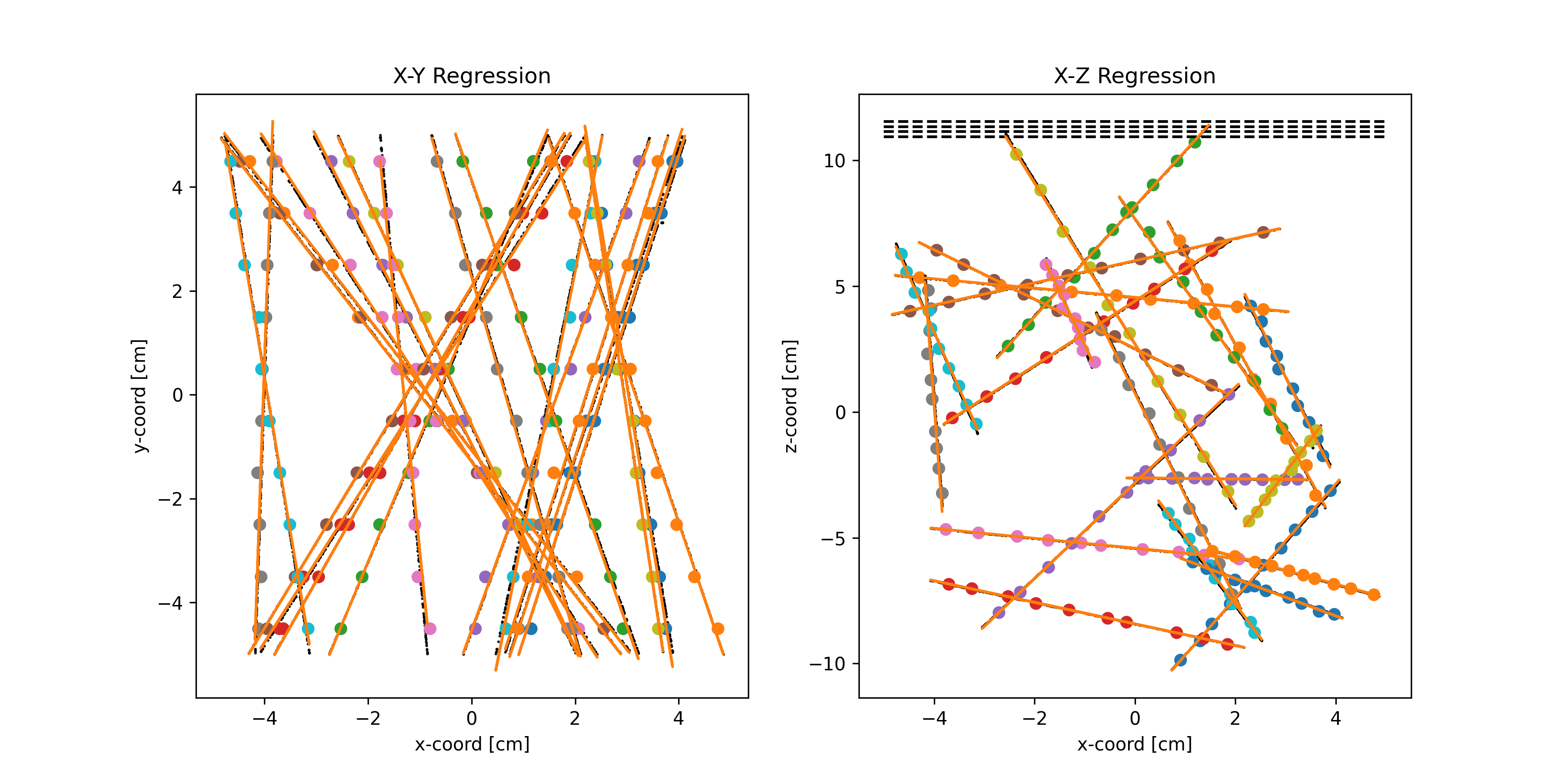}
  \end{center}
  \caption{Reconstructed tracks (dashed lines mark GEM positions).}
  \label{reg}
\end{figure}

Track reconstruction depends on the zigzag pad geometry. Figure~\ref{zigzag} compares three designs: the current prototype, a version with twice the pad count, and one with steeper angles. The readout consists of 10 rows, each with 25 pads. Figure~\ref{reg} shows reconstructed tracks for the prototype design. Table~\ref{tab:results2} presents residuals in y and z. Since this is a simulation, direct comparisons with true particle trajectories are possible. Increasing pad count improves z-resolution, while y-resolution is limited by the number of rows. Future work will optimize the zigzag design for maximal spatial resolution.

\vspace{-20pt} 

\vspace{-15pt} 

\vspace{-15pt} 
\begin{table}[H]
\caption{Reconstructed track residuals (all values in cm)}\label{tab:results2}
\centering
\renewcommand{\arraystretch}{1.2}
\begin{tabular}{cccc}
    \hline
    & Prototype & Double Pads & Double Angle \\ 
    \hline
    Number of pads & 250 & 510 & 250 \\ 
    Estimated residuals y & 0.005 $\pm$ 0.153 & 0.008 $\pm$ 0.143 & 0.002 $\pm$ 0.177 \\
    Estimated residuals z & 0.000 $\pm$ 0.099 & 0.000 $\pm$ 0.074 & 0.000 $\pm$ 0.060 \\ 
    Real Path residuals y & 0.007 $\pm$ 0.042 & 0.004 $\pm$ 0.038 & 0.009 $\pm$ 0.060 \\
    Real Path residuals z & 0.002 $\pm$ 0.027 & 0.003 $\pm$ 0.014 & 0.005 $\pm$ 0.017 \\
    \hline
\end{tabular}
\end{table}

\FloatBarrier
\section{Readout Simulation}
\label{sec:readout}

The readout simulation framework models detector readout channels under experimental conditions, assessing their influence on data acquisition from the Hadronic Range Detector (HDR), Lead Glass EM Calorimeter (LEC), and TPC. It provides a structured approach to simulate signal acquisition, incorporating key physical and electronic processes relevant to the HIBEAM-NNBAR experiment. A similar framework is being developed by Luna et al.~\citep{luna2024}, with this work expanding upon comparable methodologies.

\begin{figure}[htb!]
    \centering
    \includegraphics[width=0.6\linewidth]{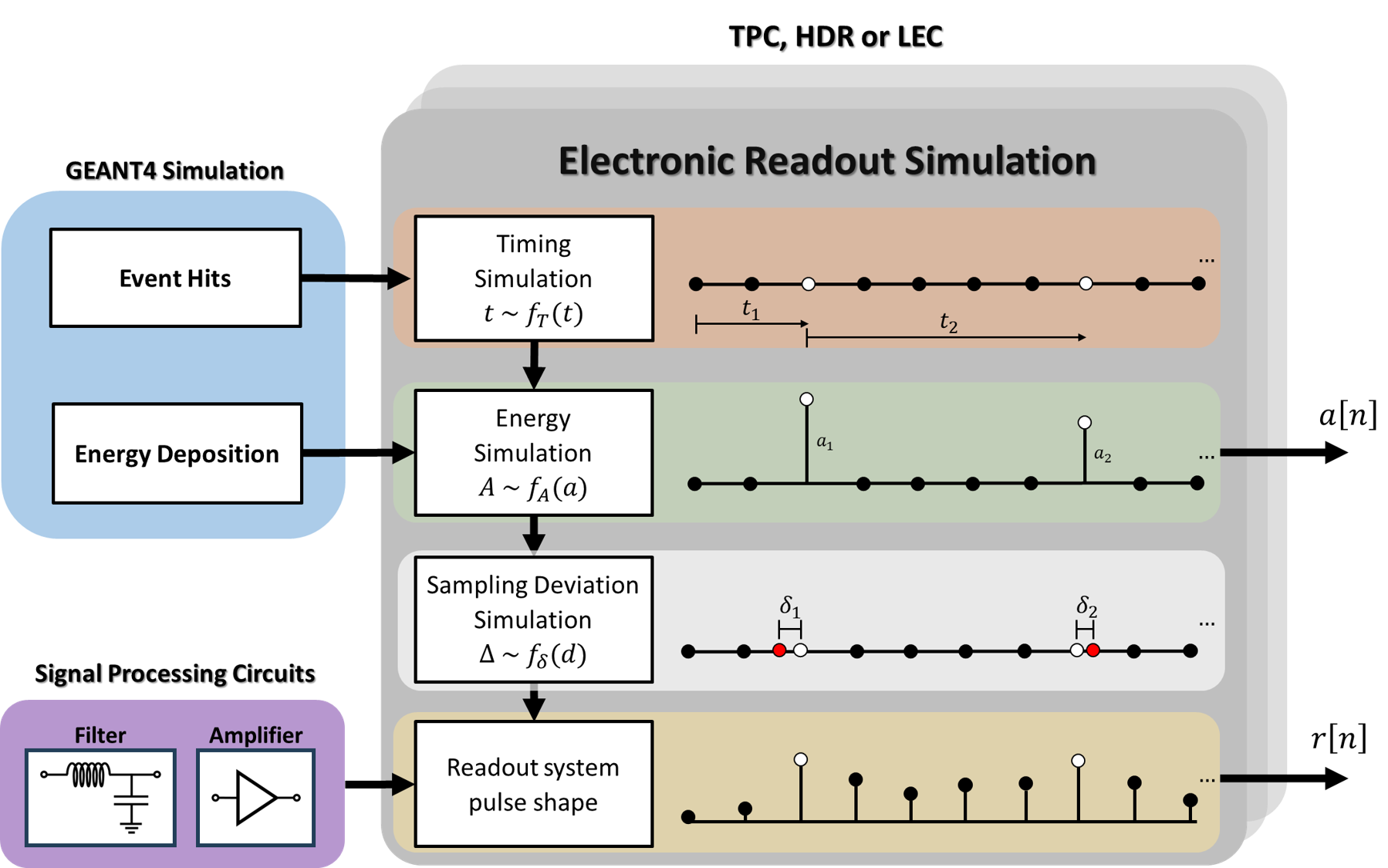}
    \caption{Flowchart of the simulation framework for TPC, HDR, or LEC readout channels.}
    \label{fig:SimFrameWork}
\end{figure}

Figure~\ref{fig:SimFrameWork} presents a flowchart outlining the framework, which ensures consistency between simulated and experimental conditions. It enables detailed investigations into noise propagation and signal distortions.

\vspace{-5pt} %

The framework consists of four main steps: \textbf{Event Timing}, modeled using a Poisson distribution to capture the stochastic nature of particle interactions; \textbf{Energy Distribution}, simulated via Geant4 to provide realistic event-by-event deposition profiles; \textbf{Readout Electronics Modeling}, which incorporates pulse shaping, amplification, and digitization effects; and \textbf{Electronic Noise}, added as a Gaussian-distributed component to simulate signal distortions.

The signal generation process follows a mathematical model where event occurrences are governed by a Poisson-distributed random variable \( t \sim f_\tau(t) \), defining a Bernoulli-distributed detection function \( p(t) \), which, in discrete time with sample period \( T_s \), is represented as \( p[n] \). The deposited energy amplitude, represented as \( A[n] \sim f_A(a) \), leads to the energy signal \( a[n] = p[n] \cdot A[n] \). 

Readout electronics shape the signal via convolution with the impulse response \( h[n] \), producing \( r[n] = a[n] * h[n] + e[n] \), where \( e[n] \) is Gaussian noise.

\begin{figure}[htb!]
    \centering
    \includegraphics[width=0.8\linewidth]{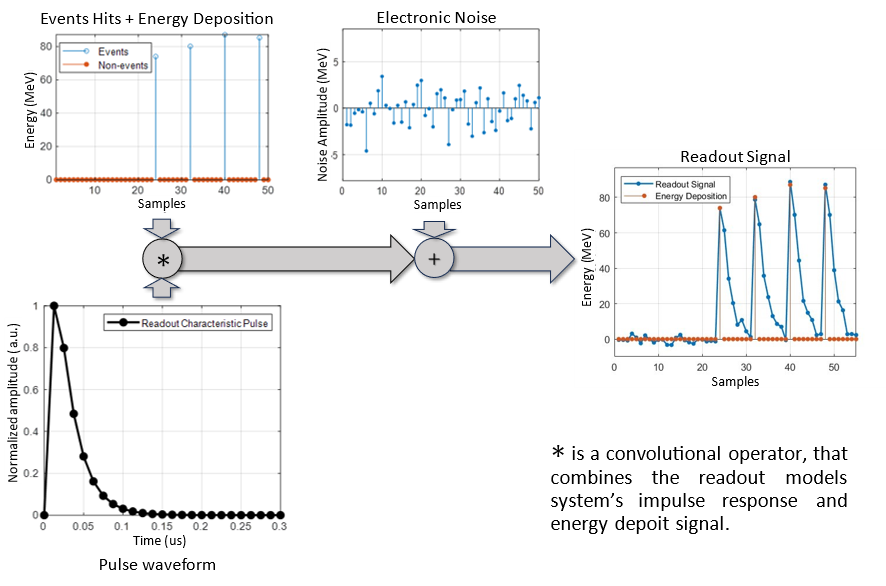}
    \caption{Results of simulated signal operations in the framework.}
    \label{fig:SimResult}
\end{figure}

Figure~\ref{fig:SimResult} shows an example of simulated signals over a few samples, assuming a 20 MHz sampling rate and an exponential decay pulse with a 0.15~$\mu$s time constant.

This simulation framework provides a robust environment for studying noise propagation and distortions in the readout system. It ensures alignment between simulated signals and experimental data, aiding in the evaluation of their impact on higher-level physics analyses, ensuring the reliability of the detector system.

\section{Summary}  
\label{summary}
This paper discusses advancements in computing and simulation techniques for the HIBEAM-NNBAR experiment at the European Spallation Source. Key topics include machine learning applications for event selection, fast parametric simulations to streamline detector studies, and detailed simulations of the time projection chamber to enhance spatial resolution. Additionally, efforts to model readout electronics are presented, addressing noise propagation and signal distortions to improve data acquisition. 

\section{Acknowledgments} 
The authors acknowledge project grant support from Vetenskapsrådet, the Olle Engkvist Foundation, and STINT - The Swedish Foundation for International Cooperation in Research and Higher Education. We also thank the Brazilian funding agencies FAPERJ, CAPES, and CNPq for their support. Detector simulation computations were enabled by resources provided by LUNARC - The Center for Scientific and Technical Computing at Lund University.

\bibliography{chep20201_bib} 
\newpage

\end{document}